\documentclass[3p,times,twocolumn]{elsarticle}

\usepackage[english]{babel}

\usepackage{ecrc}

\volume{00}

\firstpage{1}

\journalname{Nuclear Physics B Proceedings Supplement}

\runauth{E. Czerwi\'nski on behalf of KLOE and KLOE--2 collaborations}

\jid{nuphbp}

\jnltitlelogo{Nuclear Physics B Proceedings Supplement}

\usepackage{amssymb}
\usepackage[figuresright]{rotating}

\begin{document}

\begin{frontmatter}

%% Title, authors and addresses

%% use the tnoteref command within \title for footnotes;
%% use the tnotetext command for the associated footnote;
%% use the fnref command within \author or \address for footnotes;
%% use the fntext command for the associated footnote;
%% use the corref command within \author for corresponding author footnotes;
%% use the cortext command for the associated footnote;
%% use the ead command for the email address,
%% and the form \ead[url] for the home page:
%%
%% \title{Title\tnoteref{label1}}
%% \tnotetext[label1]{}
%% \author{Name\corref{cor1}\fnref{label2}}
%% \ead{email address}
%% \ead[url]{home page}
%% \fntext[label2]{}
%% \cortext[cor1]{}
%% \address{Address\fnref{label3}}
%% \fntext[label3]{}

\dochead{}
%% Use \dochead if there is an article header, e.g. \dochead{Short communication}

\title{KLOE results in kaon physics and prospects for KLOE--2}

%% use optional labels to link authors explicitly to addresses:
%% \author[label1,label2]{<author name>}
%% \address[label1]{<address>}
%% \address[label2]{<address>}

\author{E. Czerwi\'nski on behalf of KLOE and KLOE--2 collaborations\corref{collaborations}}
\cortext[collaborations]{
The KLOE collaboration:
\mbox{F.~Ambrosino,}
\mbox{A.~Antonelli,}
\mbox{M.~Antonelli,}
\mbox{F.~Archilli,}
\mbox{I.~Balwierz,}
\mbox{G.~Bencivenni,}
\mbox{C.~Bini,}
\mbox{C.~Bloise,}
\mbox{S.~Bocchetta,}
\mbox{F.~Bossi,}
\mbox{P.~Branchini,}
\mbox{G.~Capon,}
\mbox{T.~Capussela,}
\mbox{F.~Ceradini,}
\mbox{P.~Ciambrone,}
\mbox{E.~Czerwi\'nski,}
\mbox{E.~De~Lucia,}
\mbox{A.~De~Santis,}
\mbox{P.~De~Simone,}
\mbox{G.~De~Zorzi,}
\mbox{A.~Denig,}
\mbox{A.~Di~Domenico,}
\mbox{C.~Di~Donato,}
\mbox{B.~Di~Micco,}
\mbox{M.~Dreucci,}
\mbox{G.~Felici,}
\mbox{S.~Fiore,}
\mbox{P.~Franzini,}
\mbox{C.~Gatti,}
\mbox{P.~Gauzzi,}
\mbox{S.~Giovannella,}
\mbox{E.~Graziani,}
\mbox{M.~Jacewicz,}
\mbox{J.~Lee-Franzini,}
\mbox{M.~Martemianov,}
\mbox{M.~Martini,}
\mbox{P.~Massarotti,}
\mbox{S.~Meola,}
\mbox{S.~Miscetti,}
\mbox{G.~Morello,}
\mbox{M.~Moulson,}
\mbox{S.~M\"uller,}
\mbox{M.~Napolitano,}
\mbox{F.~Nguyen,}
\mbox{M.~Palutan,}
\mbox{A.~Passeri,}
\mbox{V.~Patera,}
\mbox{I.~Prado~Longhi,}
\mbox{P.~Santangelo,}
\mbox{B.~Sciascia,}
\mbox{M.~Silarski,}
\mbox{T.~Spadaro,}
\mbox{C.~Taccini,}
\mbox{L.~Tortora,}
\mbox{G.~Venanzoni,}
\mbox{R.~Versaci,}
\mbox{G.~Xu,}
\mbox{J.~Zdebik,}
and, as members of the KLOE--2 collaboration:
\mbox{D.~Babusci,}
\mbox{D.~Badoni,}
\mbox{V.~Bocci,}
\mbox{A.~Budano,}
\mbox{S.~A.~Bulychjev,}
\mbox{L.~Caldeira~Balkest\aa hl,}
\mbox{P.~Campana,}
\mbox{E.~Dan\'e,}
\mbox{G.~De Robertis,}
\mbox{D.~Domenici,}
\mbox{O.~Erriquez,}
\mbox{G.~Fanizzi,}
\mbox{G.~Giardina,}
\mbox{F.~Gonnella,}
\mbox{F.~Happacher,}
\mbox{B.~H\"oistad,}
\mbox{L.~Iafolla,}
\mbox{E.~Iarocci,}
\mbox{T.~Johansson,}
\mbox{A.~Kowalewska,}
\mbox{V.~Kulikov,}
\mbox{A.~Kupsc,}
\mbox{F.~Loddo,}
\mbox{G.~Mandaglio,}
\mbox{M.~Mascolo,}
\mbox{M.~Matsyuk,}
\mbox{R.~Messi,}
\mbox{D.~Moricciani,}
\mbox{P.~Moskal,}
\mbox{A.~Ranieri,}
\mbox{C.~F.~Redmer,}
\mbox{I.~Sarra,}
\mbox{M.~Schioppa,}
\mbox{A.~Sciubba,}
\mbox{W.~Wi\'slicki,}
\mbox{M.~Wolke}
}

\address{Institute of Physics, Jagiellonian University, Cracow, Poland}

\begin{abstract}
%% Text of abstract
The $\phi$-factory DA$\Phi$NE
offers a possibility to select pure kaon beams, charged and neutral ones. 
In particular, neutral kaons from $\phi\to K_{S}K_{L}$
are produced in pairs and the detection of a $K_S$ ($K_L$) tags the presence of a $K_L$ ($K_S$).
This allows to perform precise measurements of kaon properties by means of KLOE detector.
Another advantage of a $\phi$-factory consists in fact
that the neutral kaon pairs are produced in a pure quantum state ($J^{PC}=1^{--}$), which allows
to investigate {\em CP} and {\em CPT} symmetries via quantum interference effects,
as well as the basic principles of quantum mechanics.

A review of the most recent results of the KLOE experiment at DA$\Phi$NE using pure kaon beams or
via quantum interferometry is presented together with prospects for kaon physics
at KLOE--2.
\end{abstract}


\end{frontmatter}

%% main text
%%%%%%%%%%%%%%%%%%%%%%%%%%%%%%%%%%%%%%%%%%%%%%%%%%%%%%%%%%%%%%%%%%%%%%%%%%%%%%%
\section{KLOE experiment at DA$\Phi$NE collider}
\begin{figure}[!h]
\centering
\includegraphics[width=0.45\textwidth]{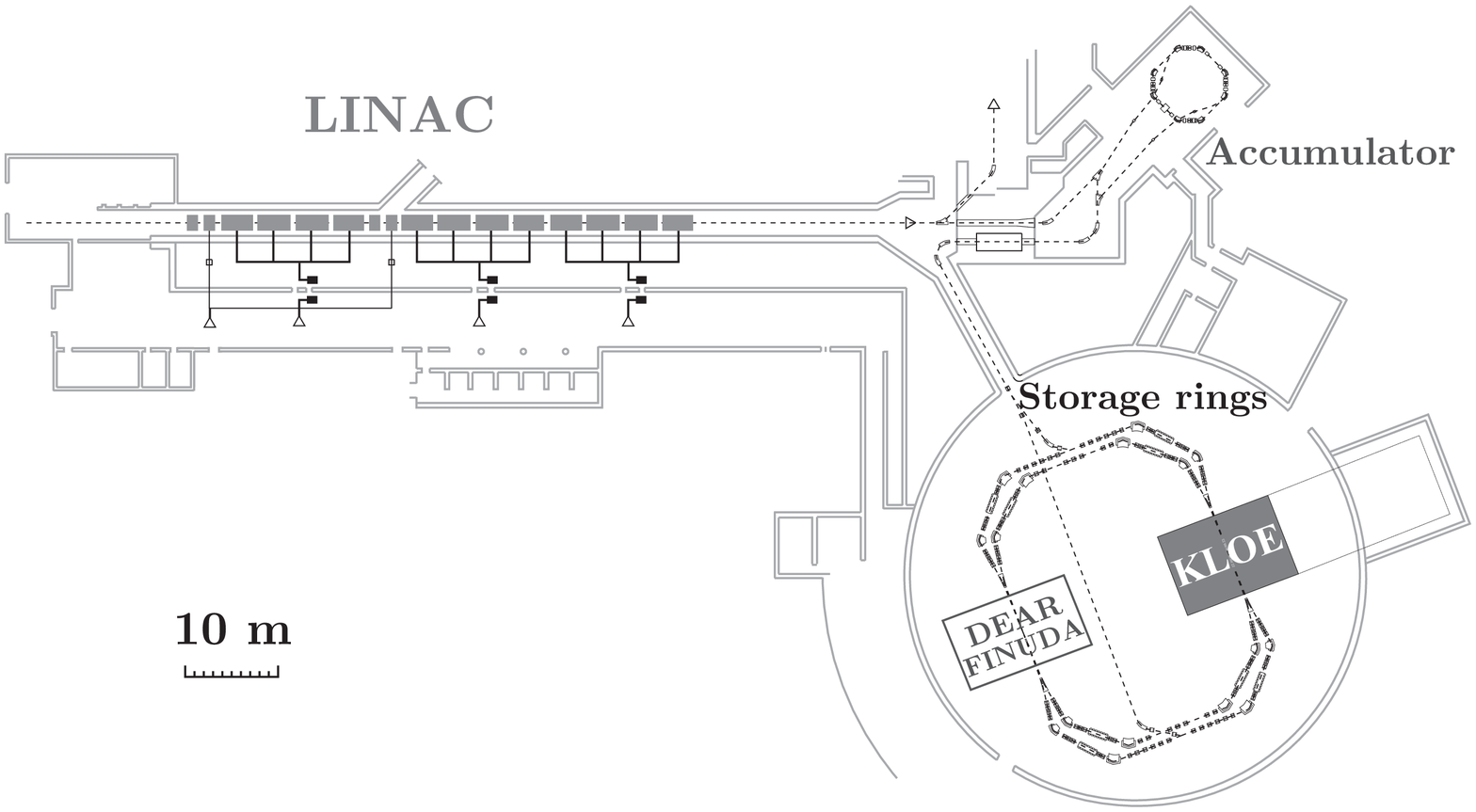}
\caption{Scheme of the DA$\Phi$NE complex. A position of KLOE detector at one of two interaction points
is also presented. Figure taken from~\cite{Franzini:2006aa}.}
\label{dafne}
\end{figure}

The e$^+$e$^-$ collider and $\phi$-factory
DA$\Phi$NE is placed in National Laboratory in Frascati (LNF-INFN, Italy).
Electrons and positrons are accelerated in linac, then stored and cooled in an accumulator and
finally transferred in bunches into separated storage rings with
two interaction points~\cite{Vignola:1996mt}. A schematic view
of DA$\Phi$NE complex is presented in Fig.~\ref{dafne}.
The collider was designed to operate at the peak of $\phi$ resonance ($\sqrt{s}=m_{\phi}\approx1019~MeV$)
producing $\phi$ mesons almost at rest ($\beta_{\phi}\approx0.015$)
since electrons and positrons collide with small transverse momenta.
A $\phi$ meson decays mostly into kaon pairs (49\% into $K^{+}K^{-}$ and 34\% into $K_{S}K_{L}$),
which makes a $\phi$-factory the natural place for kaon physics stu\-dies.

\begin{figure}[!h]
\centering
\includegraphics[width=0.30\textwidth]{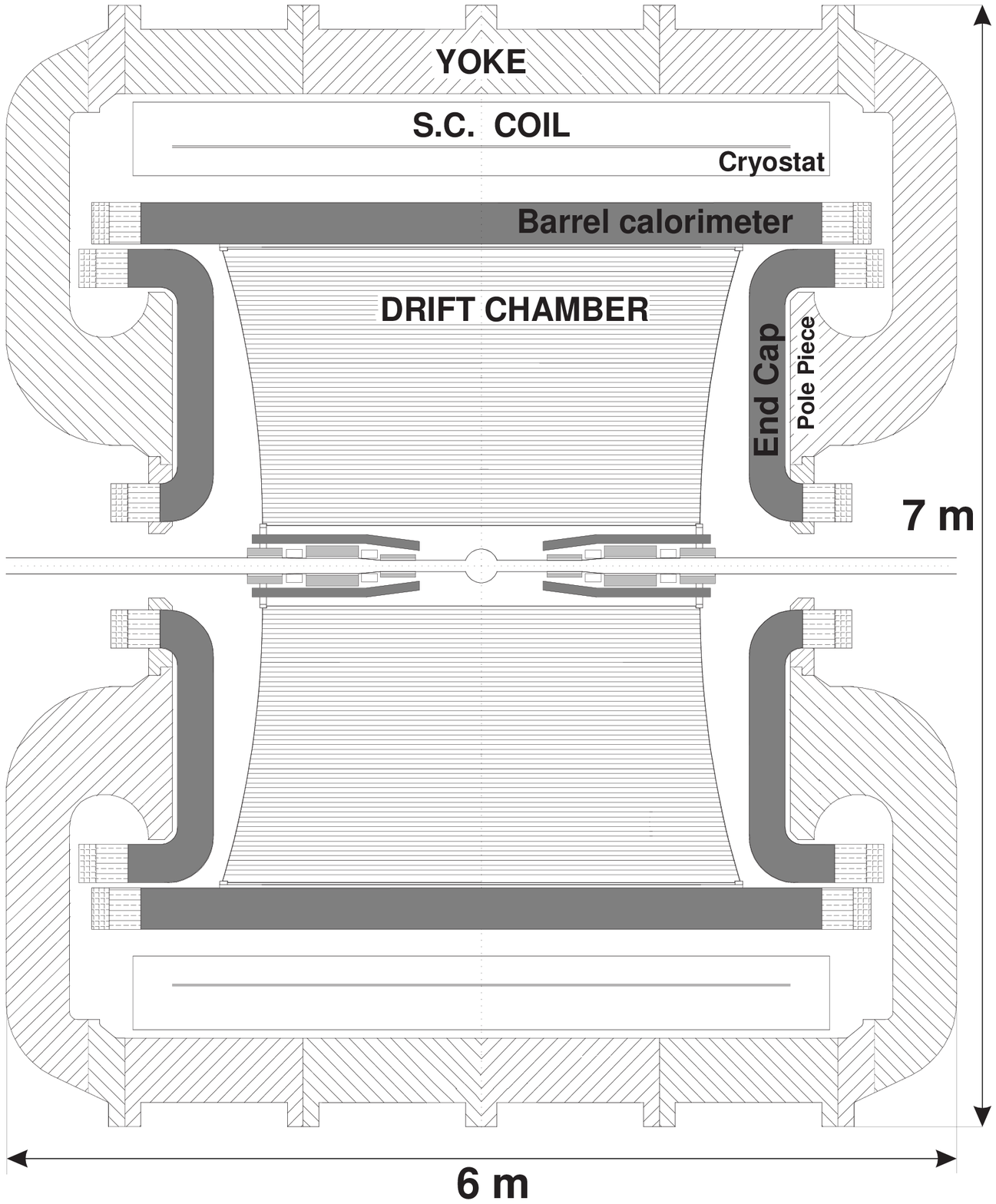}
\caption{KLOE detector surrounded by superconducting coil. Collision point of electrons and positrons is
in the spherical beam-pipe in the center of the detector. Figure adapted from~\cite{Franzini:2006aa}.}
\label{kloe}
\end{figure}

During period from 2001 to 2006 KLOE has collected 2.5~fb$^{-1}$ of integrated luminosity, which
corresponds to about $6.6\times10^9$ kaon pairs~\cite{bib:Rivista}.
The detector itself consists of two main components: a $\sim$3.3~m long cylindrical drift chamber~\cite{dc}
with radius
$\sim$4~m surrounded by an electromagnetic calorimeter~\cite{calo}. Both sub-detectors are inserted in a superconducting
coil which produces an axial magnetic field of 0.52~T parallel to the beam axis. Fig.~\ref{kloe} shows
a schematic view of KLOE detector.
The KLOE drift chamber was designed to detect a sizable fraction of $K_{L}$ decays
(mean decay path of $K_{L}$ meson is $\sim$3.4~m).
The chamber is filled with a mixture of helium and isobutan (90\% and 10\%, respectively)
and has about 52000 wires arranged in cylindrical layers with alternate stereo angles.
Based on the reconstruction of charged track curvature it allows for a fractional
momentum accuracy of $\sigma_{p}/p\approx0.5\%$. The resolution of vertex reconstruction
is about 1~mm, while overall spatial accuracy is below 2~mm.
An electromagnetic calorimeter consists of a barrel and side detectors ({\em endcaps}) providing almost 4$\pi$ coverage
of solid angle. Each module of calorimeter is read out on both sides by set of photomultipliers.
It is build of stack of 1~mm scintillating fiber layers glued between grooved lead foils.
The obtained accuracy of energy and time measurement are $\sigma_{E}=5.7\%/\sqrt{E[GeV]}$ and
$\sigma(t)=54ps/\sqrt{E[GeV]}\oplus100ps$, respectively. 
Determination of the hit position in the plane transverse to the fiber direction is based
on the analysis of signal amplitude distribution and the resolution is about 1~cm, while accuracy of longitudinal
coordinate due to excellent time resolution amounts to $\sigma_{z}=1.2cm/\sqrt{E[GeV]}$.

Since at KLOE kaons are produced in pairs from $\phi$ decay, reconstruction of $K_S$ decay close
to interaction region (clean selection of $K_{S}\to\pi^{+}\pi^{-}$, BR=69\%) allows for tagging of
$K_L$ presence. Having that and taking into account size of the detector itself makes KLOE
an excellent place for $K_L$ decay measurements. However, what is even more important, detection
of $K_L$ hit in calorimeter module tags the presence of $K_S$. This method makes KLOE an unique place to
study pure $K_S$ beams. It is used for measurement of $BR(K_{S}\to\pi^{0}\pi^{0}\pi^{0})$ described
in Section~\ref{br}.
It is also possible to study decays of both kaons in single event. Since both of them are produced
in a pure quantum state ($J^{PC}=1^{--}$), it is possible to study quantum interference effects
for reaction $K_{S}K_{L}\to\pi^{+}\pi^{-}\pi^{+}\pi^{-}$, as it is
shown in Section~\ref{interf}.

Recently KLOE--2, a successor of KLOE, has started data taking campaign in order to
extend KLOE physics program~\cite{epjC}.
Description of applied and planned
upgrades together with physics prospect is presented in Section~\ref{kloe2}.
%%%%%%%%%%%%%%%%%%%%%%%%%%%%%%%%%%%%%%%%%%%%%%%%%%%%%%%%%%%%%%%%%%%%%%%%%%%%%%%
\section{Preliminary result of $BR(K_{S}\to\pi^{0}\pi^{0}\pi^{0})$ measurement}
\label{br}
\begin{figure}[!h]
\centering
\includegraphics[width=0.30\textwidth]{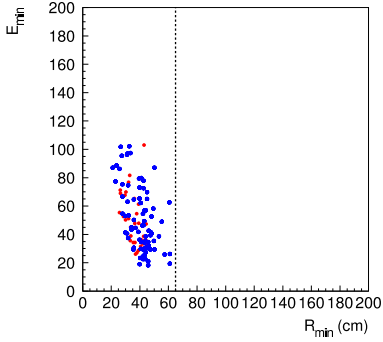}
\caption{The distribution of minimal energy of the cluster versus minimal
distance (R$_{min}$) between clusters in the event. The dashed line corresponds to the used $R_{min}$ cut.}
\label{brfig}
\end{figure}

The decay $K_{S}\to\pi^{0}\pi^{0}\pi^{0}$ has not been observed so far. 
Up to now the best limit of branching ratio is $1.2\cdot10^{-7}$~\cite{Matteo}, which is
still about two orders of magnitude larger than theoretical predictions based on
Standard Model ($BR_{SM}\sim$1.9$\cdot10^{-9}$). Assuming the {\em CPT} invariance the described process allows for
investigation of direct {\em CP} violation. At KLOE presence of $K_S$ is tagged by
$K_L$ interaction with electromagnetic calorimeter ({\em $K_L$-crash}). Additional condition to select
$K_{S}\to\pi^{0}\pi^{0}\pi^{0}$ decay candidates are presence of six neutral clusters (caused by
gamma pairs from $\pi^{0}$ decays) and no charged tracks originating from interaction point.
There are two main components for background reaction. Events of $K_{S}\to\pi^{0}\pi^{0}$
where two additional clusters are from splitting or accidental processes and
$K_{L}\to\pi^{0}\pi^{0}\pi^{0}$, $K_{S}\to\pi^{+}\pi^{-}$ events.
In the second case the $K_{L}$ decay close to the interaction point mimics the $K_{S}$ decay, while
charged products of $K_{S}$ due to interaction with quadrupoles simulate $K_L$-crash signal.
As a first step for background reduction a kinematic fit was performed with 11 constrains of
energy and momentum conservation, the kaon mass and velocity of six photons.
Cut on obtained $\chi^2$ values effectively reduce the background
from false $K_L$-crash events without loose of signal events.
For the rejection of events with splitted and/or accidental clusters
a correlation between $\chi^{2}_{3\pi}$ and $\chi^{2}_{2\pi}$ variables was used. 
Both variables are evaluated with
the most favorable cluster pairing in each case~\cite{Matteo}.
The quadratic sum of residuals between the $\pi^0$ mass
and the invariant masses of three photon pairs formed from the six clusters is $\chi^{2}_{3\pi}$,
while $\chi^{2}_{2\pi}$ is based on the invariant masses of two photon pairs and 
energy and momentum conservation for $\phi \to K_{S}K_L$ reaction, where $K_S\to\pi^0\pi^0$.
In addition, to improve the quality of the photon selection, a cut on 
$\Delta E~=~(m_{\Phi}/2 - \sum E_{\gamma_{i}})/\sigma_{E}$ was applied, where
$\gamma_i$ stands for the i--$th$ photon from  four chosen in the $\chi^{2}_{2\pi}$
estimator and $\sigma_E$ is the appropriate resolution. For two background clusters case
$\Delta E$ is expected to be 0, while $\Delta E~\approx~m_{\pi^0}/\sigma_E$ for $K_S \to 3\pi^0$.
Final cut was also applied on the minimal distance between photon clusters to refine rejection of events
with splitted clusters. Result of this cut is shown in Fig.~\ref{brfig}.

After preliminary cuts zero candidate events were obtained and zero events from Monte Carlo were observed.
Data sample is based
on 1.7 fb$^{-1}$ integrated luminosity, while effective statistics of Monte Carlo is two times higher than data sample.
This results in a new preliminary upper limit on branching ratio $BR(K_S \to 3\pi^0) < 2.9 \cdot 10^{-8}$
(corresponding to $|\eta_{3\pi^{0}}|=\big|\frac{A(K_S\to3\pi^0)}{A(K_L\to3\pi^0)}\big|<0.009$), which
is almost an order of magnitude better then the best limit obtained so far.
%%%%%%%%%%%%%%%%%%%%%%%%%%%%%%%%%%%%%%%%%%%%%%%%%%%%%%%%%%%%%%%%%%%%%%%%%%%%%%%
\section{Interference with $K_{S}K_{L}\to\pi^{+}\pi^{-}\pi^{+}\pi^{-}$}
\label{interf}
\begin{figure}[!h]
\centering
\includegraphics[width=0.22\textwidth]{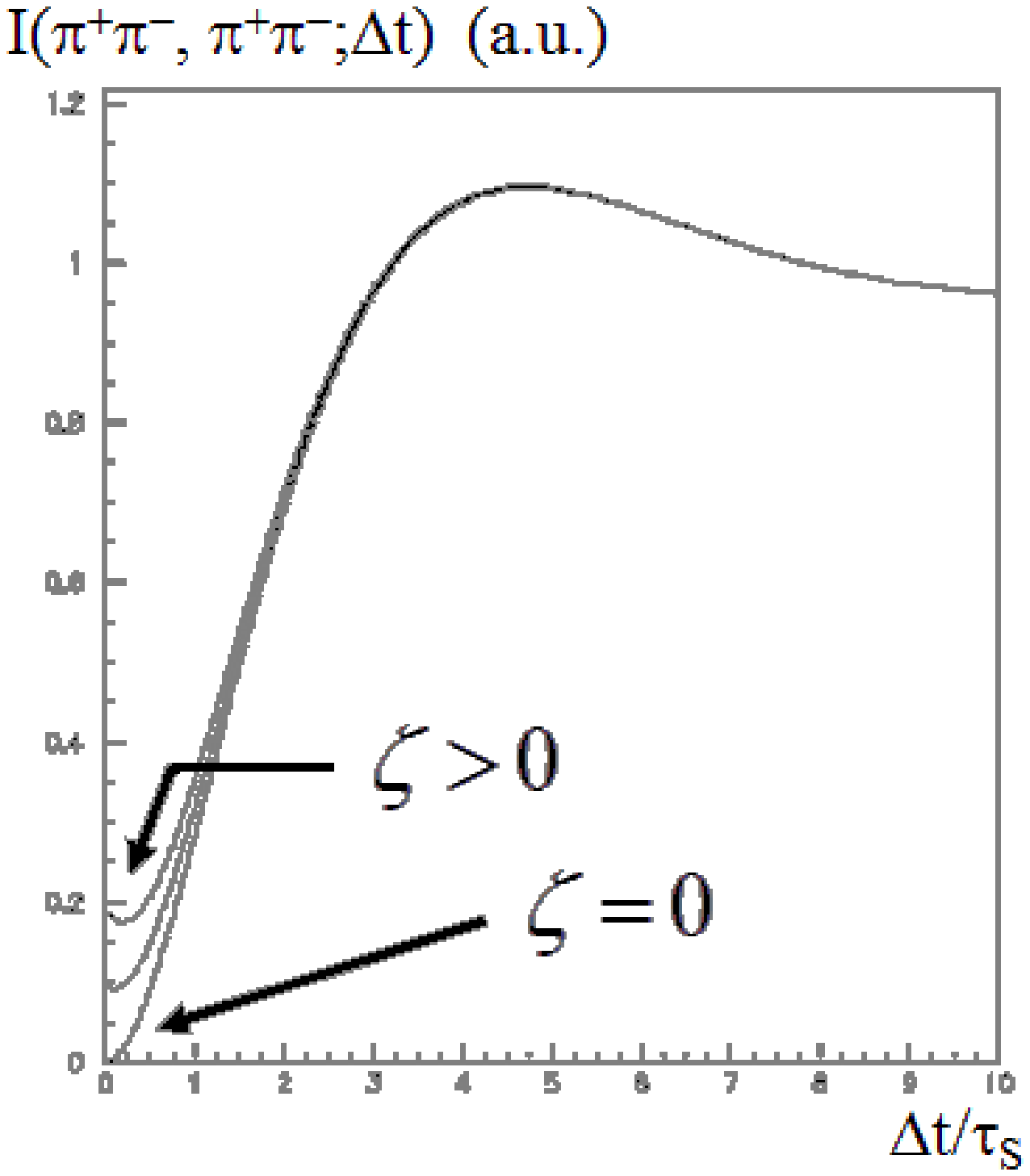}
\includegraphics[width=0.23\textwidth]{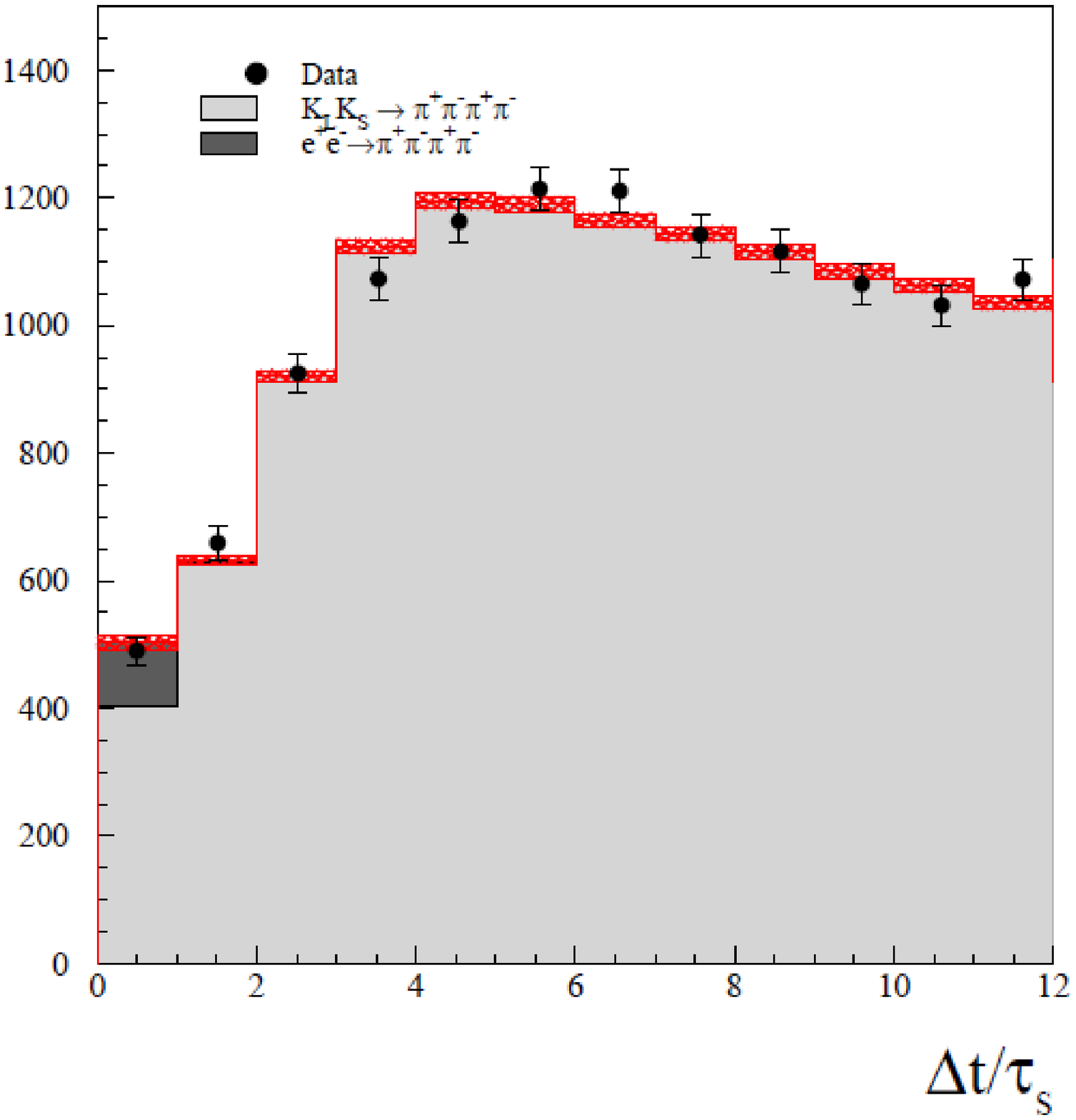}
\caption{{\bf Left:} The $I(\pi^+\pi^-, \pi^+\pi^-,\Delta t)$ distribution for quantum mechanic case ($\zeta=0$),
and for two greater than zero values of decoherence parameter. The biggest discrepancy is for $\Delta t$ close to 0.
The figure is taken from~\cite{handbook}.
{\bf Right: }Points denote experimental results, while fitted
histogram shows results of the Monte Carlo simulation. The bin size
corresponds to the time resolution $\sigma(\Delta t)\approx\tau_S$~\cite{DiDomenico:2009zza}.}
\label{interffig}
\end{figure}

At KLOE neutral kaons are produced in an entangled states from decay of $\phi$ meson with $J^{PC}=1^{--}$
\begin{eqnarray}
|i\rangle=\frac{1}{\sqrt{2}}\big\{|K^{0}\rangle|\bar{K}^{0}\rangle-|\bar{K}^{0}\rangle|K^{0}\rangle\big\}=\nonumber\\
\frac{N}{\sqrt{2}}\big\{|K_{S}\rangle|K_{L}\rangle-|K_{L}\rangle|K_{S}\rangle\big\}~,
\label{interfeq1}
\end{eqnarray}
where $N=(1+|\epsilon|^{2})/(1-\epsilon^{2})\simeq1$ is a normalization factor, with $\epsilon$ as the {\em CP} violating
parameter in the mixing~\cite{handbook}.
If one considers the case in which both kaons decay into identical final states, for example
$K_S\to\pi^+\pi^-$ and $K_L\to\pi^+\pi^-$,
the decay rate of the system is proportional to:
\begin{eqnarray}
I(\pi^+\pi^-, \pi^+\pi^-,\Delta t)~\propto~e^{-\Gamma_L \Delta t}~+~e^{-\Gamma_S \Delta t} \nonumber \\
-2e^{-\frac{\Gamma_L~+~\Gamma_S}{2}\Delta t}~cos(\Delta m \Delta t)~,
\label{interfeq2}
\end{eqnarray}
where $\Delta m$ is the mass difference between $K_L$ and $K_S$ and
$\Delta t$ is the time difference between the decay of both kaons.
This allows to study phenomena with increased accuracy due to interference pattern
$2e^{-\frac{\Gamma_L~+~\Gamma_S}{2}\Delta t}~cos(\Delta m \Delta t)$.
Eq.~\ref{interfeq2} implies that both kaons cannot decay at the same time,
and it is an example of correlation of the type
pointed out for the first time by Einstein, Podolsky and Rosen~\cite{epr}.
The initial state after $\phi$ meson decay spontaneously factorizes to an equal weight mixture of states
$|K_{S}\rangle|K_{L}\rangle$ or $|K_{L}\rangle|K_{S}\rangle$ causing decoherence.
The decoherence parameter $\zeta$, which measures the amount of deviation from prediction of
quantum mechanics, can be introduced in the following way:
\begin{eqnarray}
I(\pi^+\pi^-, \pi^+\pi^-,\Delta t)~\propto~e^{-\Gamma_L \Delta t}~+~e^{-\Gamma_S \Delta t} \nonumber \\
-2(1-\zeta_{SL})e^{-\frac{\Gamma_L~+~\Gamma_S}{2}\Delta t}~cos(\Delta m \Delta t)~,
\label{interfeq3}
\end{eqnarray}
where value $\zeta=0$ corresponds to the usual quantum mechanics case, while $\zeta=1$ to the total
decoherence (and different $\zeta$ values to intermediate situations between these two),
as it is shown at the left plot in Fig.~\ref{interffig}.
In general $\zeta$ depends on the basis in which the initial state is expressed.
At KLOE test of decoherence parameter was based on data analysis of $\approx$1.5~fb$^{-1}$~\cite{DiDomenico:2009zza}.
Selection of the signal events $\phi \to K_SK_L \to \pi^+\pi^-\pi^+\pi^-$ requires two vertexes, each with two
opposite-curvature tracks inside the drift chamber, with an invariant mass and total momentum compatible
with the two neutral kaon decays.
The experimental points were fitted with
Eq.~\ref{interfeq3} modified with parameters expressing decoherence in different models described
in~\cite{handbook,DiDomenico:2009zza}.
The fit was performed taking into account resolution and detection efficiency,
the background from coherent and incoherent $K_S$
regeneration on the beam pipe wall,
and the small contamination from the non-resonant $e^+e^- \to \pi^+\pi^-\pi^+\pi^-$ channel.
The determined experimental distribution of the $\phi \to K_SK_L \to \pi^+\pi^-\pi^+\pi^-$
intensity as a function of the absolute value of $\Delta t$ is shown on the right plot in Fig.~\ref{interffig}.
The results:
\begin{eqnarray}
\zeta_{SL}=(0.3\pm1.8_{stat}\pm0.6_{syst})\cdot10^{-2}~,\nonumber\\
\zeta_{0\bar{0}}=(1.4\pm9.5_{stat}\pm3.8_{syst})\cdot10^{-7}~,
\label{interfeq4}
\end{eqnarray}
show no deviation from quantum mechanics~\cite{DiDomenico:2009zza}.
%%%%%%%%%%%%%%%%%%%%%%%%%%%%%%%%%%%%%%%%%%%%%%%%%%%%%%%%%%%%%%%%%%%%%%%%%%%%%%%
\section{KLOE--2}
\label{kloe2}
The physics program of KLOE was related to the good accuracy of reconstruction of $K_L$ decays in large
fiducial volume, while at KLOE--2 an increased interest will be focused on the physics close
to the interaction point (IP) as rare $K_S$ decays, $K_{S}-K_L$ interference,
multi-lepton events, as well as $\eta$, $\eta'$ and $K^\pm$ decays.
More details about whole KLOE--2 physics program can be found in Ref.~\cite{epjC}.

The new interaction region of electron and positron beams based on the Crabbed Waist compensation of the beam-beam
interaction together with large Piwinski angle and small beam sizes
at the crossing point was successfully commissioned in 2011~\cite{pantaleo}.
This modification allows to increase delivered luminosity (with unchanged beam current) by a factor
of three with respect to the performance reached before the upgrade.
During next 3-4 years of data taking with the KLOE--2 detector it is planed to collect
$\sim$20~fb$^{-1}$ of integrated luminosity.

The detector itself was upgraded with two pairs of small angle tagging devices~\cite{taggers} to detect
low (Low Energy Tagger - LET)~\cite{bib:let} and high (High Energy Tagger - HET)
energy $e^{+}e^{-}$ originated from $e^{+}e^{-}\to e^{+}e^{-}X$ reactions. This tagger system
will be used for $\gamma\gamma$ physics studies.
First commissioning run of KLOE--2 has started in 2011.
After collection of $\sim$5~fb$^{-1}$ the second phase of upgrade will start.
In this step a light-material Inner Tracker detector
based on the Cylindrical GEM technology
will be installed in the region between the beam pipe and the drift chamber to
improve charged vertex reconstruction and to increase the acceptance for low transverse momentum tracks~\cite{it1,it2}.
Crystal calorimeters (CCALT) will cover
the low polar $\theta$ angle
region to increase acceptance for very forward electrons and photons down to 8$^\circ$~\cite{CCALT}.
A new tile calorimeter (QCALT) will be
used for the detection of photons coming from $K_L$ decays in the drift chamber~\cite{QCALT}.

The tagging system is made of two different detectors which are already installed and ready for the data taking.
Near the interaction region inside KLOE the small calorimeters (LET) consisting of LYSO crystals read out
by silicon photomultipliers are placed.
They will be used for measurements of electrons and positrons from $\gamma\gamma$ interaction
within energy range from 160 to 230~MeV with an accuracy of $\sigma_E$~$\sim$~10\%~\cite{taggers}.
At the distance of 11~m from the interaction point in the bending section of DA$\Phi$NE, the HET detectors are placed.
They provide the measurement of the displacement of the scattered $e^{+}$ or $e^{-}$ with respect to the main
orbit.
These position detectors consist of 30 small BC418 scintillators 3x3x5 mm$^3$ each and provides a spatial resolution of 2~mm
(corresponding to momentum resolution of $\sim$1~MeV) for $e^{+}/e^{-}$  with energy higher than 400~MeV.

The main part of KLOE--2 physics program~\cite{epjC} is concentrated in $K_S$, $\eta$
and charged kaon decays as well as kaon interferometry.
These events are produced close to the interaction point (IP),
requiring an optimization of the detection for low momentum tracks coming
from the IP.
This is the purpose of installation of the Inner Tracker detector.
It is based on a novel technology of fully cylindrical GEM (Gas Electron Multiplier) detectors~\cite{it1,it2}.
Each of its four concentric layers provides 2 coordinates, while the third is determined
by the known radius of the layer.
Each layer is a triple-GEM chamber with cathode and anode made of thin polyimide foils.
The innermost layer will be placed 15~cm from the beam line, which corresponds to 20~$\tau_S$
in order not to spoil the $K_{L}K_{S}$
interference, while the outermost layer is located close to the internal wall of the Drift Chamber.

\begin{figure}[!h]
\centering
\includegraphics[width=0.30\textwidth]{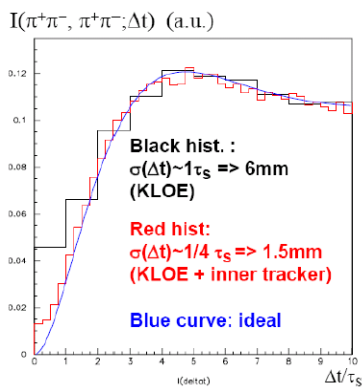}
\caption{Comparison of the $I(\pi^+\pi^-, \pi^+\pi^-,\Delta t)$ distribution obtained with KLOE resolution
and after Inner Tracker insertion based on the Monte Carlo simulation.}
\label{kloe2fig}
\end{figure}

The comparison between results obtained with present spatial resolution and after installation of
Inner Tracker for decoherence studies (discussed in Section~\ref{interf}) is presented in Fig.~\ref{kloe2fig}.
An improved sensitivity on decoherence parameters of about one order of magnitude is expected with an
integrated luminosity of $\sim$20~fb$^{-1}$ and the use of Inner Tracker.

In order to improve measurements of the rare kaon, $\eta$, and $\eta'$ decays two additional calorimeters 
QCALT and CCALT will be installed. QCALT composed by a sampling of 5~layers of 5~mm thick scintillator
tiles alternated with 3.5~mm thick tungsten plates will constitute a 1~m long dodecagonal structure covering the region
of the new quadrupoles inside the KLOE detector.
The active part of each plane is divided into twenty tiles of about 5x5 cm$^2$ area with 1~mm
diameter WLS fibers embedded in circular grooves. Silicon photomultipliers of 1 mm$^2$ area are used for readout
each fiber~\cite{QCALT}.
In addition, crystal calorimeters (CCALT) shaped as two small barrels
of LYSO crystals  will cover the low polar angle region very close to the IP to increase acceptance
for very forward photons. Readout system with silicon photomultipliers will be used
aiming to achieve a timing resolution between 300 and 500~ps for 20~MeV
photons~\cite{CCALT}.
%%%%%%%%%%%%%%%%%%%%%%%%%%%%%%%%%%%%%%%%%%%%%%%%%%%%%%%%%%%%%%%%%%%%%%%%%%%%%%%
\section{Summary}
KLOE experiment achieved already several significant and precise results
in kaon physics~\cite{bib:Rivista}, due to the unique possibility of producing pure $K_L$, $K_S$ and $K^\pm$ beams.
Moreover, there are still ongoing investigation, like search of $K_S\to3\pi^0$ decay presented here.
The success of the DA{$\Phi$}NE upgrade motivated the start-up of a new experiment KLOE--2
which aims to complete and extend the KLOE physics program.
LET and HET detectors have been already installed and the first phase of new data taking campaign has started.
After collection of about 5~fb$^{-1}$ of data new set of detectors will be installed (Inner Tracker, QCALT and CCALT
calorimeters) for precise measurements of rare decays of kaon, $\eta$ and $\eta'$, as well
as CP and CPT tests and kaon interferometry studies.
The expected total integrated luminosity collected by KLOE--2 should be about 20 fb$^{-1}$.
The accuracy in the measurements is expected in most cases to be improved by about one order
of magnitude~\cite{epjC}.
%%%%%%%%%%%%%%%%%%%%%%%%%%%%%%%%%%%%%%%%%%%%%%%%%%%%%%%%%%%%%%%%%%%%%%%%%%%%%%%
%% For references without a BibTeX database:


\section*{References}
\begin{thebibliography}{00}
\bibitem{Vignola:1996mt}
  The $\phi$-factory Study Group {\em Proposal for a $\phi$-factory}, LNF-90/031 (R) (1991).
\bibitem{Franzini:2006aa}
  P.~Franzini and M.~Moulson,
  %``The Physics of DAFNE and KLOE,''
  Ann.\ Rev.\ Nucl.\ Part.\ Sci.\  {\bf 56} (2006) 207.
  %[hep-ex/0606033].
  %%CITATION = HEP-EX/0606033;%%
\bibitem{bib:Rivista}
  F.~Bossi,E.~De Lucia,J.~Lee-Franzini, S.~Miscetti, M.~Palutan and KLOE Collaboration,
  %{\it Precision Kaon and Hadron Physics with KLOE},
  Rivista del Nuovo Cimento Vol.31, N.10 (2008).
\bibitem{dc}
  M. Adinolfi, {\it et al.},
  %The KLOE drift chamber,
  Nucl. Instrum. Meth. A  {\bf 461} (2001) 25-28.
\bibitem{calo}
  M.~Adinolfi {\it et al.},
  %``The KLOE electromagnetic calorimeter,''
  Nucl.\ Instrum.\ Meth.\ A {\bf 482} (2002) 364.
  %%CITATION = NUIMA,A482,364;%%
\bibitem{epjC}
  G. Amelino-Camelia {\it et al.},
  %Physics with the KLOE-2 experiment at the upgraded DA$\Phi$NE,
  Eur. Phys. J. C {\bf 68} (2010) 619-681.
\bibitem{Matteo}
  F. Ambrosino {\it et al.},
  %A direct search for the CP-violating decay $K_S \to 3\pi^0$ with the KLOE detector at DA$\Phi$NE,
  Phys. Lett. B {\bf 619} (2005) 61-70.
\bibitem{handbook}
  {\em Handbook on neutral kaon interferometry at a Phi-factory}, editor: A.~Di~Domenico, Frascati Physics Series
  {\bf 43} (2007).
\bibitem{epr}
  A.~Einstein, B.~Podolsky, N.~Rosen, Physical Review {\bf 47} (1935) 777.
\bibitem{DiDomenico:2009zza}
  A.~Di Domenico {\it et al.}  [KLOE Collaboration],
  %``Search for CPT violation and decoherence effects in the neutral kaon system,''
  J.\ Phys.\ Conf.\ Ser.\  {\bf 171} (2009) 012008.
  %%CITATION = 00462,171,012008;%%
\bibitem{pantaleo}
  M Zobov {\it et al.},
  %Test of "`Crab-Waist"' Collisions at the DA$\Phi$NE $\phi$--Factory,
  Phys. Rev. Lett. {\bf 104} (2010) 174801-174806.
\bibitem{taggers}
  F.~Archilli {\it et al.},
  Nucl. Instr. \& Meth. A {\bf 617} (2010) 266.
\bibitem{bib:let}
  D.~Babusci {\it et al.},
  Nucl. Instr. \& Meth. A {\bf 617} (2010) 81.
\bibitem{it1}
  KLOE-2 Collaboration, F.~Archilli {\it et al.},
  %{\it Technical Design Report of the Inner Tracker for the KLOE-2 experiment }, arXiv:1002.2572v1 and
  LNF-10/3(P) INFN-LNF, arXiv:1002.2572
\bibitem{it2} 
  A.~Balla {\it et al.},
  Nucl. Instr. \& Meth. A {\bf 604} (2009) 23.
\bibitem{CCALT}
  M.~Cordelli {\it et al.},
  Nucl. Instr. \& Meth.  A {\bf 617} (2010) 109.
\bibitem{QCALT}
  M.~Cordelli {\it et al.},
  Nucl. Instr. \& Meth. A {\bf 617} (2010) 105.
\end{thebibliography}
\end{document}